# <arttitle> Parahydrogen enhanced zero-field nuclear magnetic resonance

<aug> T. Theis[1,2*], P. Ganssle[1,2*], G. Kervern[1,2*], S. Knappe[3], and J. Kitching[3], M. P. Ledbetter[4*], D. Budker[3,4], and A. Pines[1,2]

<aff> [1]*Materials Science Division, Lawrence Berkeley National Laboratory, Berkeley CA 94720, [2]Department of Chemistry, University of California at Berkeley, Berkeley, California 94720-3220, [3]Time and Frequency Division, National Institute of Standards and Technology, 325 Broadway, Boulder, Colorado, 80305, [4]Department of Physics, University of California at Berkeley, Berkeley, California 94720-7300, [5]Nuclear Science Division, Lawrence Berkeley National Laboratory, Berkeley CA 94720*

<aff>  * These authors contributed equally to this work

<abs> Nuclear magnetic resonance (NMR) [1,2], conventionally detected in multi-tesla magnetic fields, is a powerful analytical tool for the determination of molecular identity, structure, and function.  With the advent of prepolarization methods and alternative detection schemes using atomic magnetometers[3,4] or superconducting quantum interference devices (SQUIDs)[5], NMR in very low- (~earth's field), and even zero-field, has recently attracted considerable attention[6,7,8, 9,10,11].  Despite the use of SQUIDs or atomic magnetometers, low-field NMR typically suffers from low sensitivity compared to conventional high-field NMR.  Here we demonstrate direct detection of zero-field NMR signals generated via parahydrogen induced polarization (PHIP)[12], enabling high-resolution NMR without the use of any magnets.  The sensitivity is sufficient to observe spectra exhibiting [13]C-[1]H J-couplings in compounds with [13]C in natural abundance in a single transient.  The resulting spectra display distinct features **that have straightforward interpretation and can be used for chemical fingerprinting.**

<p> NMR in low or zero magnetic field has long been viewed as a curiosity due to the low nuclear spin polarization, poor sensitivity of inductive pickup coils at low frequencies, and the absence of molecule-specific chemical shifts. In this contribution, we show that the first two objections can be overcome, and that, despite the lack of chemical shift, significant spectroscopic information remains. The advantage of working in low- or zero magnetic-field is that it allows for narrow lines due to the high absolute field homogeneity and the correspondingly accurate determination of coupling parameters. Additionally, eliminating superconducting magnets and cryogenics facilitates the development of inexpensive and mobile sensors for chemical analysis. The development of pre-and hyperpolarization methods, and the advent of alternative methods for detection of nuclear spins have increased sensitivity dramatically, enabling chemical analysis by low-field NMR in a variety of contexts. For example, chemical shifts[7] in $^{129}Xe$ and spin-spin or $J$-couplings[8] of $^{1}H$-$^{29}Si$ and $^{1}H$-$^{19}F$ have been detected by inductive pickup coils in earth's field NMR and used for chemical analysis. Atomic magnetometers[3,4] and SQUIDs[5] are sensitive to low-frequency signals, offering dramatically improved signal-to-noise ratio in low-field NMR, and have been used to detect low-field NMR [6,9,10,11] and to acquire magnetic resonance images[13] . Following Ref. [6], we use an atomic magnetometer to detect $J$-couplings in a zero-field environment.

To produce nuclear spin polarization in zero-field NMR, we employ the technique of parahydrogen induced polarization (PHIP), whereby order from the singlet state of parahydrogen is transferred to a molecule of interest, either by hydrogenation[12,14,15,16], or through reversible chemical exchange[17]. When combined with sensitive atomic magnetometers for detection of nuclear spin magnetization, this enables NMR without any magnets. The sensitivity is sufficient to easily observe complex spectra exhibiting $^1$H-$^{13}$C J-couplings in compounds with $^{13}$C in natural abundance in just a few transients, a task that would require considerable signal averaging using only a prepolarizing magnet. While PHIP has been investigated in a variety of magnetic fields, ranging from the earth's field to high field, observation of the resulting NMR signals has always been performed in finite magnetic field. To the best of our knowledge, the work reported here represents the first direct observation of PHIP in a zero-field environment. We show that polarization can be transferred through a number of chemical bonds to remote parts of a molecule, and that zero-field spectroscopy can be used to distinguish between different isotopomers in ethylbenzene, the product of hydrogenation of styrene. The mechanism by which observable magnetization is generated from the parahydrogen derived singlet order requires only the presence of a heteronucleus, similar to the work of Ref. [18], in contrast to a more commonly observed mechanism relevant to high field, which requires chemical shift differences at the sites of the parahydrogen derived protons. Furthermore, our results are of particular interest in the context of recent work demonstrating that the lifetime of singlet polarization in low fields can considerably exceed the $T_1$ of $z$ magnetization[19,20]. These demonstrations of increased singlet lifetime relied on field cycling and high field inductive detection, and our methodology may provide for more direct observation and exploitation of these effects.

<p> Zero-field NMR spectroscopy of samples magnetized by thermal prepolarization in a permanent magnet was discussed in Refs. [6,21]. In an isotropic liquid at zero magnetic field, the only terms in the NMR Hamiltonian are the spin-spin J-couplings, $H_J = \sum \hbar J_{jk} \mathbf{I}_j \cdot \mathbf{I}_k$. In the important case of $AX_N$ systems, where both $A$ and $X$ are spin-1/2 particles, and each $X$ spin couples to $A$ with the same strength $J$, the resulting zero field J-spectra are simple and straightforward to interpret, consisting of a single line at $J$ for $AX$, a single line at $3J/2$ for $AX_2$, and two lines, one at $J$ and one at $2J$ for $AX_3$. For larger molecules, as employed in the present work, long-range couplings to additional spins lead to splitting of these lines, however, the overall positions of the resulting multiplets remain unchanged. Presently we rely on numerical spin simulations (presented in detail in the Supplementary Information) to understand the splitting pattern, however we anticipate that an approach based on perturbation theory will likely yield simple rules for interpretation of the zero field splitting pattern.

Zero-field spectroscopy using parahydrogen induced polarization differs from the case of thermal polarization in both the initial density matrix and in the method of excitation. In the case of homogeneous catalysis, the product molecule starts out with two parahydrogen derived spins in a singlet state. Averaging over random hydrogenation events and subsequent evolution under the J-coupling Hamiltonian lead to an equilibrium density matrix described by pairs of heteronuclear and homonuclear scalar spin pairs, $\rho_0 = \sum a_{jk} \mathbf{I}_j \cdot \mathbf{I}_k$, which bears no magnetic moment, and is static under the $J$-coupling Hamiltonian. Observable magnetization oscillating along the $z$ direction, to which the magnetometer is sensitive can be produced by applying a pulse of DC magnetic field $B$ in the $z$ direction. Immediately following such a pulse, the density matrix contains terms of the form $\sin \eta (I_{jx} I_{ky} - I_{jy} I_{kx})$, where $\eta = B t_p (\gamma_j - \gamma_k)$, $t_p$ is the pulse duration, and $\gamma_j$ is the gyromagnetic ratio of spin $j$. Subsequent evolution under the J-coupling Hamiltonian results in terms in the density matrix of the form $(I_{j,z} - I_{k,z}) \sin \eta \sin(J_{jk} t)$, which produces magnetization oscillating in the $z$ direction. The dependence on $\eta$ highlights the role of a heteronucleus in the symmetry breaking of the parahydrogen derived scalar order. Numerical spin simulations of the propagation of the parahydrogen derived scalar order through the molecule and the dependence of the coherence amplitude on pulse area $\eta$ for a heteronuclear spin pair with scalar order are presented in the Supplementary Information. A more detailed analysis of the polarization transfer will be presented in a forthcoming publication.

<p> The zero-field spectrometer used in this work is similar to that of Ref. [6] and is shown in Fig. 1 (a). The noise spectrum of the magnetometer is shown in (b), and the pulse sequence is shown in (c). We performed zero-field PHIP spectroscopy in hydrogenation reactions of styrene (which forms ethylbenzene) and 3-hexyne (hexene and hexane), 1-phenyl-1propyne (1-phenyl-1propene) and dimethylacetylenedicarboxlyate (dimethylmaleate). In measurements presented in the main text, parahydrogen was bubbled through the solution for ~10 s, the flow was halted, and excitation pulses of DC magnetic field were applied in the $z$ direction with $\eta=\pi/2$ for $^{13}$C and protons. The resulting $z$ magnetization was recorded by the atomic magnetometer. The rate of hydrogenation can be monitored by the signal amplitude as a function of time, as presented in the Supplementary Information. More details of the experimental setup and procedures can be found in Methods.

<p> Single shot, zero-field PHIP spectra of ethylbenzene-$\beta^{13}$C (labelled $^{13}$CH$_3$ group), and ethylbenzene-$\alpha^{13}$C (labelled $^{13}$CH$_2$ group), synthesized from labelled styrene, are shown in black in Fig. 2 (a) and (b), respectively. The ethylbenzene molecule is shown in the inset, with the blue carbon indicating the β label, and the green carbon indicating the α label. The spectrum of ethylbenzene-$\beta^{13}$C in Fig. 2(a) can be understood in terms of the discussion above, with multiplets at $^1J_{HC}$ and $2\times^1J_{HC}$, and additional lines at low frequency. Here the superscript indicates the number of bonds separating the interacting pair, and for ethylbenzene-$\beta^{13}$C, $^1J_{HC}$=126.2 Hz (Ref. [22]). Isolated lines in the complex spectrum fit to complex Lorentzians with half-width-at-half-maximum (HWHM) of about 0.1 Hz. It should be noted that this spectrum is similar to the correspondingly labelled ethanol-$\beta^{13}$C spectrum reported in Ref [6], although careful inspection reveals small splittings of some lines due to long-range (at least four-bond) homonuclear couplings to protons on the benzene ring. The blue trace shows the result of a numerical simulation accounting for eight spins, including the six spins on the ethyl part of the molecule and the two nearest protons on the benzene ring.

The simulation reproduces most of the features of the experimental spectrum quite well, including small splittings of several lines.  More details of the numerical simulation can be found in the Supplementary Information.

<p> The zero-field PHIP spectrum of ethylbenzene-$\alpha$$^{13}$C shown in Fig. 2 (b) is qualitatively similar to the zero-field spectrum of ethanol-$\alpha$$^{13}$C [6], with a multiplet at roughly 3/2× $^1J_{HC}$ ($^1J_{HC}$ = 126.2 Hz, measured in house with a 300 MHz spectrometer) and features at low frequency. Many additional lines in the spectrum indicate that long-range couplings to the protons on the benzene ring are important. Since the ethanol-$\beta$$^{13}$C spectrum does not display such complexity, the largest perturbation to the ethyl part of the molecule must be due to three-bond $^3J_{HC}$ couplings.  The green trace shows the result of numerical simulation, consisting of the six spins on the ethyl part of the molecule and the two nearest protons on the benzene ring.  Simulation again reproduces most of features of the experimental spectrum, although careful inspection shows a number of additional splittings in the experimental spectrum, indicating that couplings to more remote spins on the benzene ring not included in the simulation, are important. It is worth emphasizing that, despite the similarity of the one-bond heteronuclear $J$-couplings, the spectra associated with different isotopomers display strikingly different features, which appear in different parts of the spectrum, facilitating easy assignment of isotopomers to their respective peaks.

<p> The sensitivity of the magnetometer and the degree of parahydrogen induced polarization are sufficient to detect $J$-spectra in compounds with $^{13}$C in natural abundance.  Figure 3 shows the zero-field PHIP spectrum of ethylbenzene with $^{13}$C in natural abundance, obtained in just eight transients.  The spectrum shown here is the sum of spectra associated with the $\alpha$ and $\beta$ isotopomers shown in Fig. 2, as well as isotopomers that carry $^{13}$C is in one of four non-equivalent positions on the benzene ring. The high frequency parts of the spectrum arising from the $\alpha$ and $\beta$ isotopomers are highlighted in green and blue, respectively. The part of the signal arising from the benzene ring with a single $^{13}C$ is a multiplet centred about the one-bond coupling frequencies (typically about 156 Hz in aromatic systems), and also a multiplet in the low-frequency range. The high-frequency component is highlighted in red. Interestingly, spectra associated with the $\alpha$ or $\beta$ isotopomers do not overlap with spectra associated with isotopomers with a $^{13}C$ on the benzene ring. It is also noteworthy that if the hydrogenation is performed in high field, large chemical shift differences between protons on the benzene ring and the parahydrogen derived protons would inhibit the transfer of polarization to the benzene ring.

<p> To further illustrate the capabilities of zero-field PHIP as a method for chemical fingerprinting, we present spectra obtained from several different hydrogenation reactions, shown in Fig. 4: (a) phenylpropyne with a labelled $^{13}CH_3$ group, (b) dimethylacetylenedicarboxylic acid with $^{13}C$ in natural abundance, and (c) 3-hexyne with $^{13}C$ in natural abundance. These spectra are the result of averaging 1, 6, and 32 transients respectively. The phenylpropyne spectrum displays characteristics similar to the ethylbenzene-$\beta$ $^{13}C$ spectrum, although the phase and splitting pattern is clearly different since neither of the parahydrogen derived protons are part of the labelled group. The acetylene dimethylmaleate spectrum shown in (b) is the superposition of two different $^{13}C$ isotopomers, and can approximately be understood as follows: For a three-spin system, where one of the parahydrogen derived spins has a strong coupling to a $^{13}C$ nucleus, one can show that the spectrum consists of two lines centred around the strong coupling frequency, and an additional low frequency peak. The antiphase lines centred about 165 Hz in (b) correspond to the isotopomer where the $^{13}C$ is directly bonded to one of the parahydrogen derived spins, and is accompanied by a contribution at low frequency. The other three-spin isotopomer, where the strongest coupling to the $^{13}C$ nucleus is through two bonds, nominally gives rise to three lines at low frequency. There are some residual splittings in the low-frequency part of the spectrum, which will be subject of future investigations. The spectrum obtained in the hexyne reaction in (c) is the sum of three different $^{13}$C isotopomers. For labelled $^{13}$CH$_3$ groups, signal arises at $^1J_{HC}$ and $2\times{}^1J_{HC}$, where $^1J_{HC} \approx 125$ Hz. For labelled $^{13}$CH$_2$ groups, the contribution to the signal is centred about $3J/2$, producing signal in the range of 170 to 200 Hz. Long range couplings to other spins yield additional splitting. A more detailed discussion of these spectra and rules for assigning transitions and understanding ZF-PHIP spectra will be presented in a forthcoming publication.

Finally, we make several observations: 1) Here we operate in zero magnetic field. Working in small but finite fields on the order of 1 mG may yield additional information regarding molecular structure, albeit at the expense of additional spectral complexity.[23] 2) A common objection to low- and zero-field NMR is that spectra become complex as the number of spins increase, as exemplified by comparison of the the ethanol-α$^{13}$C spectrum reported in Ref. [6] and the ethylbenzene-α $^{13}$C obtained here. The increasing complexity of spectra with spin system size is a feature that is also encountered in standard high-field NMR, and has been successfully addressed by application of multi-pulse sequences and multidimensional spectroscopy. The theory of multiple pulse sequences for zero-field NMR has been worked out some time ago[24], and presumably, many of the techniques developed for high field could be adapted to zero-field. 3) We achieve linewidths of about 0.1 Hz. For $^{13}$C-H J-coupled systems, the dispersion in signal is about 300 Hz, so roughly 1500 lines can fit in a spectrum without overlapping. This is similar to what may be achieved in a 400 MHz spectrometer if we assume proton chemical shifts ranging over 6 ppm and proton linewidths of about 0.5 Hz. 4) The sensitivity of the magnetometer used in this work was about 0.15 nG/Hz$^{1/2}$ using a vapour cell with a volume of 10 mm$^3$. Sensitivities about 2 orders of magnitude better have been achieved in larger vapour cells[25], which will enable measurements on larger samples with much lower concentration.

<p> In conclusion, we have demonstrated NMR without the use of any magnets by using parahydrogen induced polarization and a high sensitivity atomic magnetometer with a microfabricated vapour cell. The mechanism by which the symmetry of the singlet states is broken in zero field relies only upon the presence of heteronuclear J-coupling and not chemical shifts, in contrast to many experiments performed in high field. Hydrogen-carbon J-couplings through at least three bonds, and hydrogen-hydrogen couplings through four bonds are observed. We also observe that polarization is naturally transferred through several bonds to remote parts of the molecule. This can be contrasted with *in-situ* hydrogenation in high field, where chemical shifts larger than *J*-couplings prevent efficient polarization transfer without the use of auxiliary RF pulses. Sensitivity is sufficient to perform J-spectroscopy on samples with $^{13}C$ in natural abundance with very little signal averaging. The resulting spectra, while exhibiting a large overall number of lines, can easily be divided into different parts, which can directly be assigned to different isotopomers of the molecule at hand. While our technique may appear limited to molecules to which hydrogen can be added, recent advances using iridium complex catalysts enable polarization of molecules without hydrogenation[26], significantly expanding the scope of applicability of zero-field PHIP. Since the development of zero-field NMR is still at an early stage it is not possible to fully gauge its competitiveness with high field NMR or portable lower resolution versions thereof, but it clearly has potential to become a low cost, portable method for chemical analysis.

<meth1ttl> **Methods**

<meth1hd> **Experimental setup**

The zero-field spectrometer is similar to that of Ref. [6] and is shown schematically in Fig. 1 (a) An atomic magnetometer, consisting of a Rb vapour cell and two lasers for optical pumping and probing, operates in the spin-exchange relaxation-free[4] regime.

The cell is placed inside a set of magnetic shields (not shown), and residual magnetic fields are zeroed to within ≈1 μG. The vapour cell has dimensions 5 mm × 2 mm × 1 mm, contains $^{87}$Rb and 1300 torr of $N_2$ buffer gas mm, and was microfabricated using lithographic patterning and etching techniques. The cell is heated to 210 ℃ via an electric heating element wound around an aluminium-nitride spool. The sensitivity of the magnetometer is about 0.15 nG/Hz$^{1/2}$ above 120 Hz, and the bandwidth is in excess of 400 Hz. A set of coils can be used to apply sharp, ≈1 G DC pulses in arbitrary directions to excite NMR coherences, and a separate set of coils (not shown) controls the ambient magnetic field inside the shields. Mixtures of catalyst, solvent, and substrate could be brought into proximity of the atomic magnetometer via a glass sample tube. The sample was maintained at 80℃ by flowing air through a jacket surrounding the glass tube. In experiments at lower temperature (not presented here), we found that there was some nonuniform broadening of spectra, presumably due to the presence of catalyst in solid form. Parahydrogen was bubbled through the solution via 0.8 mm inner-diameter tube for several seconds at a pressure of about 70 PSI and flow rate of about 120 standard cc/min. Bubbling was halted prior to application of excitation pulses and signal acquisition. Data were acquired with sampling rate of 2 kS/s. In acquiring the spectrum of styrene with natural-abundance $^{13}$C, the phase of the excitation pulses was cycled with respect to that of the 60 Hz line frequency in order to reduce the line noise and its harmonics.

<meth1hd> **Production of parahydrogen**

Parahydrogen was produced at 29 K by flowing hydrogen gas through a bed of iron oxide catalyst in a setup similar to that described in Ref. [27] and then stored in an aluminium canister at room temperature and initial pressure of 150 PSI. Conversion of hydrogen to parahydrogen was about 95%, and storage lifetime was in excess of one week.

**<meth1hd> Sample preparation**

Isotopically labeled styrene was obtained from Cambridge Isotope Labs.  Natural-abundance styrene and Wilkinson's catalyst[28] were obtained from Sigma-Aldrich.

Styrene hydrogenations were performed with 300 μL styrene and 4 mG Wilkinson's catalyst, Tris(triphenylphosphine)rhodium(I) chloride

 (CAS # 14694-95-2).  The phenylpropyne and dimethylacetylenedicarboxylate reactions were performed with 100 μL substrate in 300 ml tetrahydrofuran, catalyzed by 1,4-Bis(diphenylphosphino)butane](1,5-cyclooctadiene)rhodium(I)Tetrafluoroborate, (CAS # 79255-71-3). The hexyne reaction was performed in a solution of 50% tetrahydrofuran with 5 mL total and 30 mG Wilkinson's Catalyst.  Most of this volume does not contribute to signal since it is far from the magnetometer.

<ack> We acknowledge useful discussions with M. Levitt and magnetometer-cell fabrication help from S. Schima. Research was supported by the U.S. Department of Energy, Office of Basic Energy Sciences, Division of Materials Sciences and Engineering under Contract No. DE-AC02-05CH11231 (T. Theis, P. Ganssle, G. Kervern, A. Pines), by the National Science Foundation under award #CHE-0957655 (D. Budker and M. P. Ledbetter), by the National Institute of Standards and Technology (S. Knappe and J. Kitching), Civilian Research & Development Foundation Award # RU-C1-2915-NO-07.  This work is a partial contribution of NIST and is not subject to US copyright.

<corr> Correspondence and requests for materials should be addressed to A. Pines (pines@berkeley.edu).

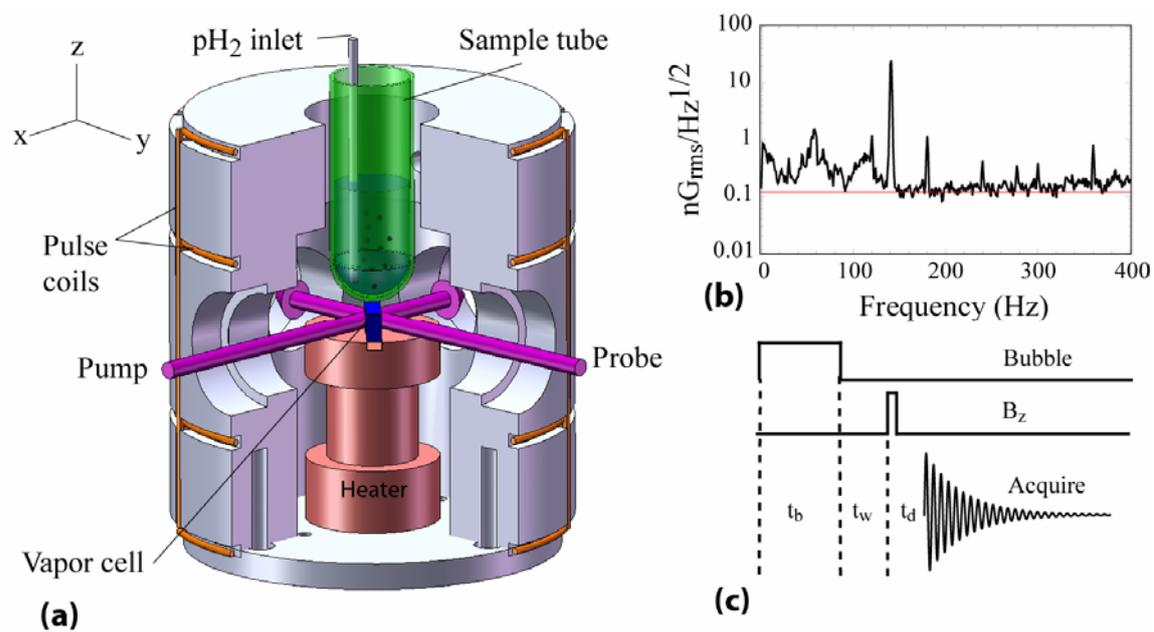

**Figure (1)**

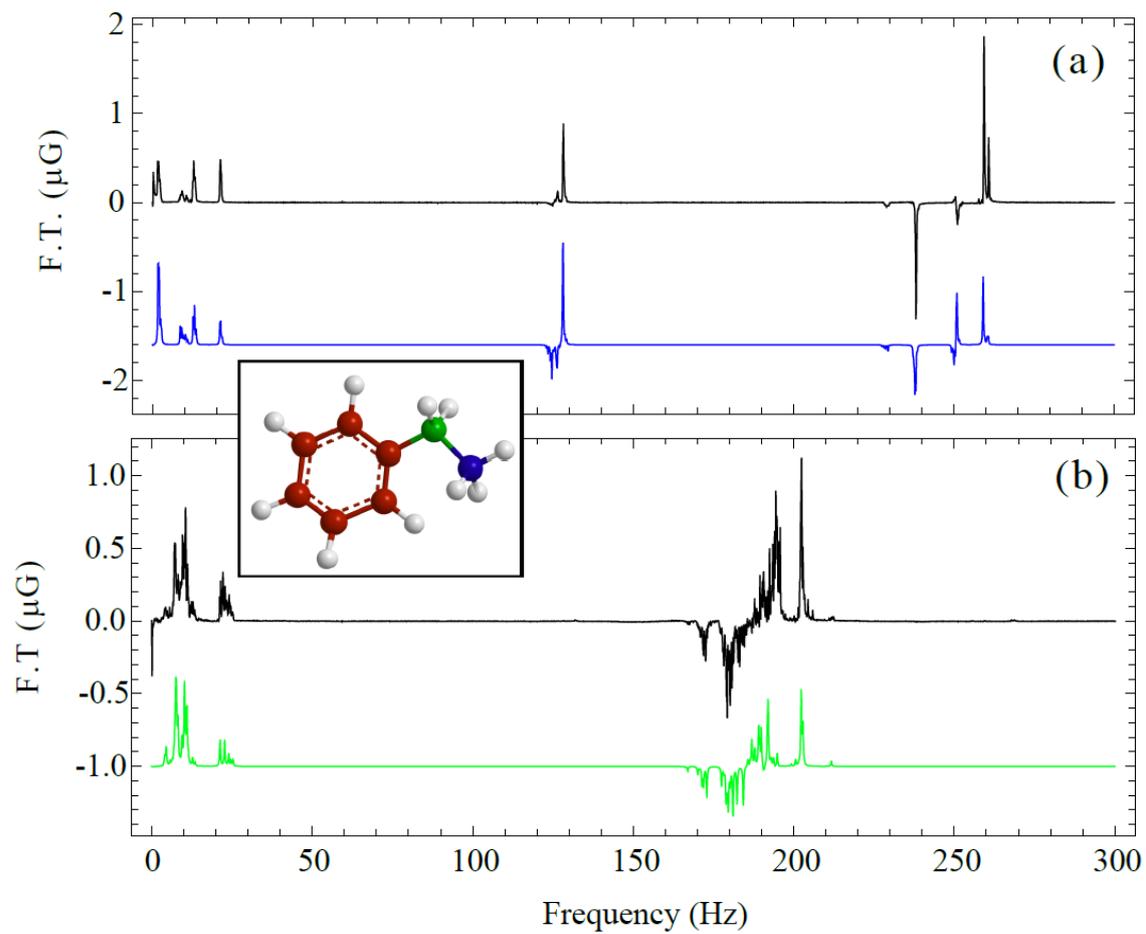

**Figure 2**

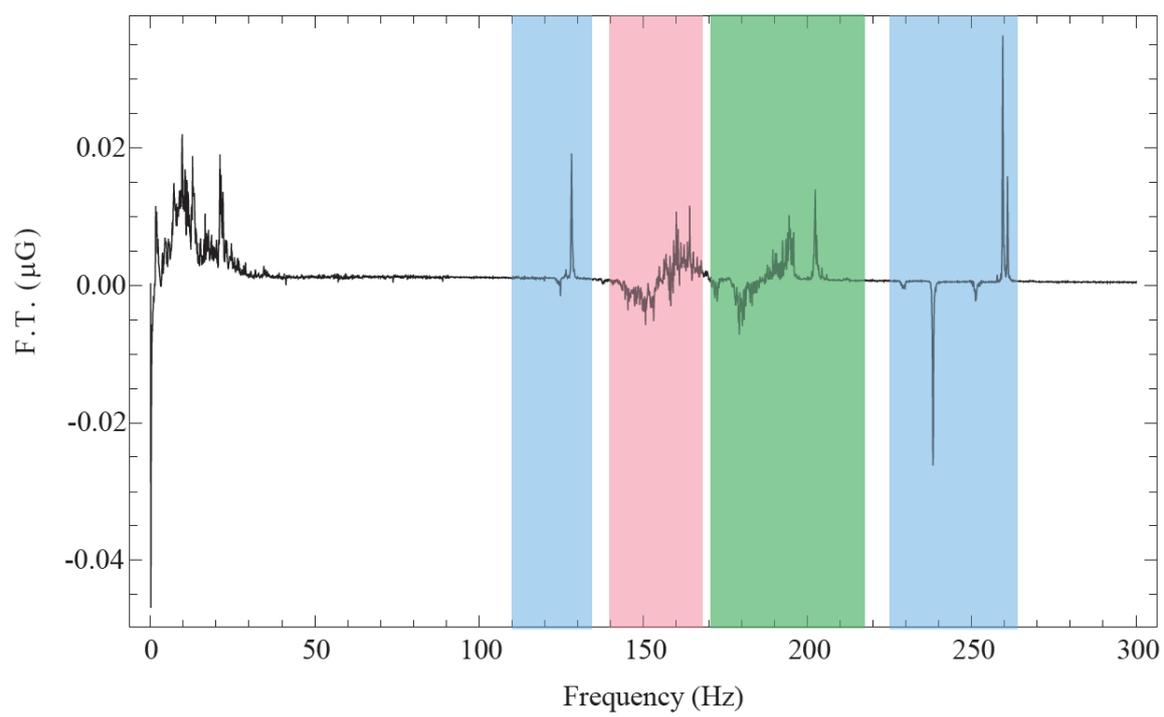

**Figure 3**

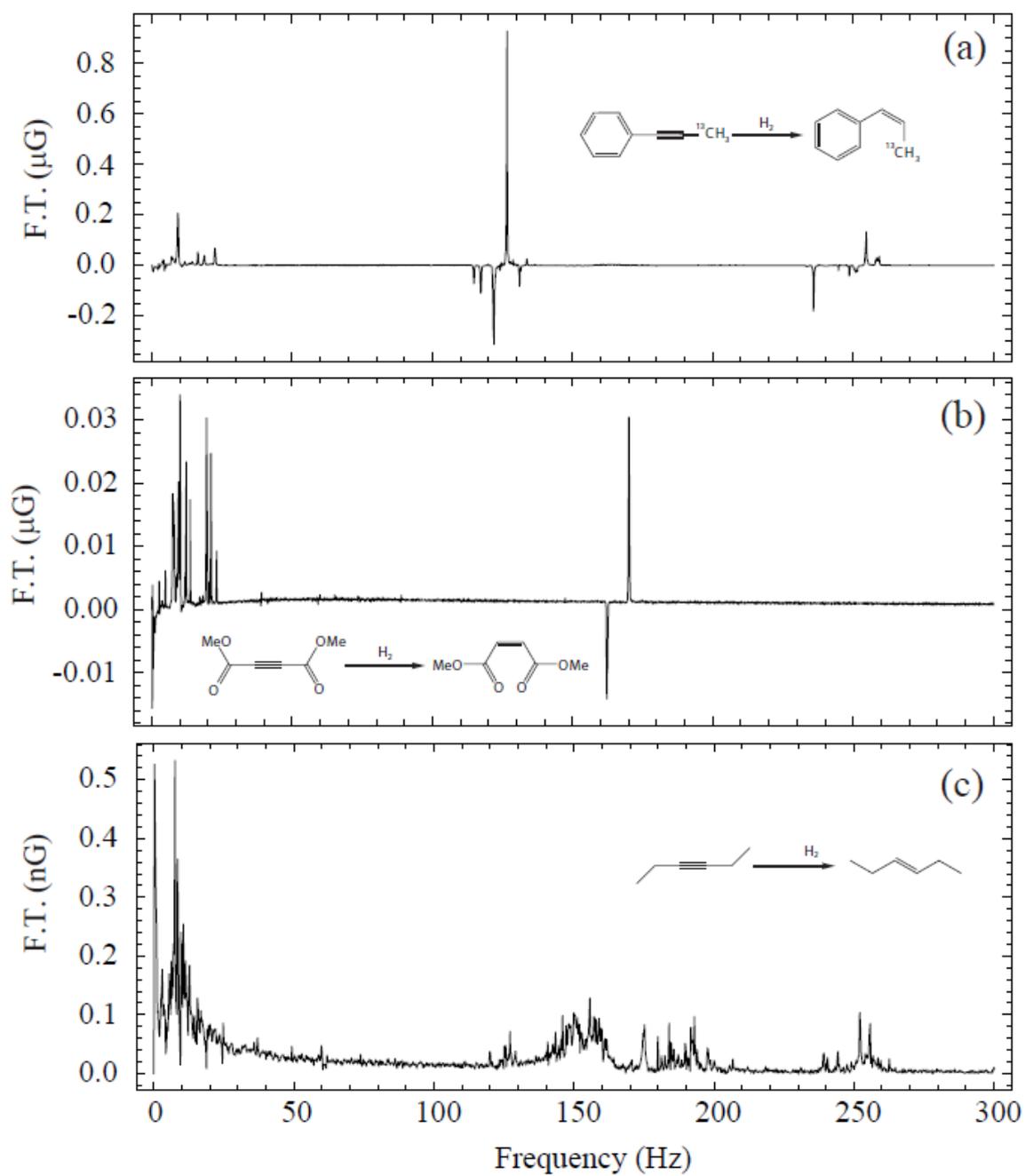

**Figure 4**

<LEGEND> Figure 1: (a) Experimental setup. A microfabricated alkali vapor cell is mounted inside a set of coils used for applying magnetic field pulses. A circularly polarized laser beam, resonant with the D1 transition of $^{87}$Rb, optically pumps the alkali vapor. A linearly polarized laser beam, tuned about 100 GHz off resonance, is used to probe the alkali spin-precession. The magnetometer is primarily sensitive to magnetic fields in the vertical ($z$) direction. A 7 mm inner-diameter glass tube contains the sample, and a 1/32" inner-diameter teflon tube is used to bubble parahydrogen through the solution. (b) Magnetic field noise spectrum. Above 100 Hz, the noise floor is about 0.15 nG/Hz$^{1/2}$. (c) Experimental sequence, showing bubbling, excitation, and data acquisition.

<LEGEND> Figure 2: Single-shot zero-field PHIP J-spectra (imaginary component) of ethylbenzene-β$^{13}$C (a) and ethylbenzene-α$^{13}$C (b), polarized via addition of parahydrogen to labelled styrene. The inset shows the ethylbenzene molecule with the β and α positions indicated by the blue and green carbons, respectively. The blue and green traces in (a) and (b) are the results of numerical simulations, described in the text.

<LEGEND> Figure 3: (a) Zero-field J-spectrum (magnitude) of ethylbenzene produced, produced via parahydrogenation of styrene with $^{13}$C in natural abundance. These data result from averaging 8 transients following a pulse of magnetic field in the $z$ direction with η≈π/2. Contributions from the α and β isotopomers are easily recognizable from the spectra shown in Fig. 2. The signal in the neighbourhood of 156 Hz is due to isotopomers with $^{13}$C on the benzene ring.

<LEGEND> Figure 4: ZF-PHIP spectra for hydrogenation of 1-phenyl-1-propyne, labelled with $^{13}C$ in the $CH_3$ group (a), acetylene dimethylcarboxylate with $^{13}C$ in natural abundance (b), 3-hexyne with $^{13}C$ in natural abundance (c). In (a) and (b), the imaginary component is presented, in (c), magnitude is presented.

---

[1] Ernst, R. R., Bodenhausen, G., A. Wokaun, *Principles of Nuclear Magnetic Resonance in One and Two Dimensions*, Oxford University Press, New York (1987).

[2] Slichter, C. P., *Principles of Magnetic Resonance, 3rd ed.* Springer-Verlag, New York (1990).

[3] Budker, D. & Romalis, M. V. Optical Magnetometry. *Nature Physics* **3**, 227 (2007).

[4] Kominis, I. K., Kornack, T. W., Allred, J. C., & Romalis, M. V. A sub-femtoTesla multichannel atomic magnetometer, *Nature*. **422**, 596 (2003).

[5] Greenberg, Y. S. Application of superconducting quantum interference devices to nuclear magnetic resonance. *Reviews of Modern Physics* **70**, 175-222 (1998).

[6] Ledbetter, M. P. *et al*. Optical detection of NMR J-spectra at zero magnetic field. *J. Magn. Res.* **199**, 25-29, (2009).

[7] Appelt, S., Häsing, F. W., Kühn, H., Perlo, J., & Blümich, B. Mobile high resolution xenon nuclear magnetic resonance spectroscopy in the Earth's magnetic field. *Phys. Rev. Lett.* **94**, 197601-197604 (2005).

[8] Appelt, S., Kühn, H., Häsing, F. W., & Blümich, B. Chemical analysis by ultrahigh-resolution nuclear magnetic resonance in the Earths magnetic field. *Nat. Phys.* **2**, 105-109, (2006).

[9] Savukov, I. M. & Romalis, M. V. NMR detection with an atomic magnetometer. *Phys. Rev. Lett.* **94**, 123001 (2005).

[10] Ledbetter, M. P. *et al*. Zero-field remote detection of NMR with a microfabricated atomic magnetometer. *Proc. Nat. Acad. Sci. (USA)* **105**, 2286-2290, (2008).

[11] McDermott, R. et al. Liquid-state NMR and scalar couplings in microtesla magnetic fields. *Science* **295**, 2247-2249 (2002).

[12] Bowers, C. R. & Weitekamp, D. P. Transformation of symmetrization order to nuclear-spin magnetization by chemical-reaction and nuclear-magnetic-resonance. *Phys. Rev. Lett.* **57**, 2645-2648 (1986).

[13] Xu, S. J. *et al.* Magnetic resonance imaging with an optical atomic magnetometer. *Proceedings of the National Academy of Sciences of the United States of America* **103**, 12668-12671, doi:10.1073/pnas.0605396103 (2006).

[14] Natterer, J. & Bargon, J. Parahydrogen induced polarization. *Prog in Nucl. Magn. Res. Spec.* **31**, 293-315 (1997).

[15] Bowers, C.R. Sensitivity enhancement utilizing parahydrogen. Encyplodia of Nuclear Magnetic Resonance **9**, 750-770 Eds. D.M. Grant and R.K. Harris (2002).

[16] Canet, D. *et al.* Para-hydrogen enrichment and hyperpolarization. *Concepts in Magnetic Resonance Part A* **28A**, 321-330, doi:10.1002/cmr.a.20065 (2006).

[17] Adams *et al.* Reversible interactions with para-hydrogen enhance NMR sensitivity by polarization transfer, *Science* **323**, 1708 (2009).

[18] Aime, S., Gobetto, R., Reineri, F., Canet, D., Polarization transfer to heteronuclei: The effect of H/D substitution. The case of the AA'X and $A_2A_2'X$ spin systems. *J. Magn. Res.,* **178**, 184 (2006).

[19] Carravetta, M., Johannessen, O. G. & Levitt, M. H. Beyond the $T_1$ limit: singlet nuclear spin states in low magnetic fields. *Phys. Rev. Lett.* **92**, 153001-153004 (2004).

[20] Pileio, G., Carravetta, M. & Levitt, M. H. Extremely Low-Frequency Spectroscopy in Low-Field Nuclear Magnetic Resonance. *Phys. Rev. Lett.* **103**, 083002 (2009).

[21] Zax, D.B., Bielecki, A., Zilm, K.W., & Pines, A. Heteronuclear Zero-Field NMR, *Chem. Phys. Lett.* **106**, 550-553 (1984).

[22] Schaefer, T., Chan, W. K., Sebastian, R., Schurko, R., & Hruxka, F. E. Concerning the internal rotational barrier and the experimental and theoretical $^nJ(^{13}C, \, ^{13}C)$ and $^nJ(^1H, ^{13}C)$ in ethylbenzene-$\beta^{13}C$.

[23] Appelt, S. *et al.* Paths from weak to strong coupling in NMR, *Phys. Rev. A* **81**, 023420 (2010).

[24] Lee, C.J., Suter, D., and Pines, A., Theory of multiple-pulse NMR at low and zero fields, J. Magn. Res. **75**, 110-124 (1987).

[25] Dang, H.B., Maloof, A.C., & Romalis, M.V., Ultrahigh sensitivity magnetic field and magnetization measurements with an atomic magnetometer, *Apl. Phys. Lett.* **97**, 151110 (2010).

[26] Adams *et al.* Reversible interactions with para-hydrogen enhance NMR sensitivity by polarization transfer, *Science* **323**, 1708 (2009).

[27] Koptyug, I. V. *et al.* Para-hydrogen induced polarization in heterogeneous hydrogenation reactions. *J. Am. Chem. Soc.* **129**, 5580-5586 (2007)

[28] Osborn, J. A., Jardine, F. H., Young, J. F. & Wilkinson, G. The Preparation and Properties of Tris(triphenylphosphine)halogenorhodium(I) and Some Reactions Thereof Including Catalytic Homogeneous Hydrogenation of Olefins and Acetylenes and Their Derivatives. *J. Chem. Soc. A*, 1711–1732 (1966).